# Estimation of Characteristics of a Software Team for Implementing Effective Inspection Process through Inspection Performance Metric

T.R.Gopalakrishnan Nair
Director, Research and Industry Incubation Center (RIIC)
Professor of Computer Science
Dayananda Sagar Institutions, Kumaraswamy Layout
Bangalore-560078, India
Phone No. +91 9447793184
Fax No. + 91 80 26660789
trgnair@ieee.org
trgnair@gmail.com

Suma. V
Asst.Professor of Information Science Department
Member in Research and Industry Incubation Center
Dayananda Sagar Institutions, Kumaraswamy Layout
Bangalore-560078, India
Phone No. +91 9448305148
Fax No. +91 80 23637053
Sumavdsce@gmail.com

Nithya G. Nair
Department of Computing and Engineering
University of Ulster
Coleraine, U.K.
Nithya_g_nair@yahoo.com

**ABSTRACT**

The continued existence of any software industry depends on its capability to develop nearly zero-defect product, which is achievable through effective defect management. Inspection has proven to be one of the promising techniques of defect management. Introductions of metrics like, Depth of Inspection (DI, a process metric) and Inspection Performance Metric (IPM, a people metric) enable one to have an appropriate measurement of inspection technique. This article elucidates a mathematical approach to estimate the IPM value without depending on shop floor defect count at every time. By applying multiple linear regression models, a set of characteristic coefficients of the team is evaluated. These coefficients are calculated from the empirical projects that are sampled from the teams of product-based and service-based IT industries. A sample of three verification projects indicates a close match between the IPM values obtained from the defect count ($IPM_{dc}$) and IPM values obtained using the team coefficients using the mathematical model ($IPM_{tc}$). The IPM values observed onsite and IPM values produced by our model which are strongly matching, support the predictive capability of

IPM through team coefficients. Having finalized the value of IPM that a company should achieve for a project, it can tune the inspection influencing parameters to realize the desired quality level of IPM. Evaluation of team coefficients resolves several defect-associated issues, which are related to the management, stakeholders, outsourcing agents and customers. In addition, the coefficient vector will further aid the strategy of PSP and TSP.

1. INTRODUCTION

Any software industry has an evolutionary growth and it stabilizes with the development of quality products. One of the most critical components of quality is the realization of defect-free product in order to achieve total customer satisfaction. Figure 1. depicts the interdependency of customer satisfaction and defects. Insert Figure1. about here.

Spiewak and McRitchie (2008) state that quality is free while non-quality things cost. Hence, they suggest that the best practice of identification and fixing of process defects enable one to achieve the product quality. They further suggest the use of cost modeling tool to access the changes in cost, schedule, and quality with the changes in assumptions. They feel that the use of modeling tools for implementation of best practices reduces defects and reduces the cost to fix defects. This knowledge improves the quality and enhances the total productivity of the organization. The main intent of quality cost analysis is not to remove the cost entirely, but to ensure that the investment yields maximum benefit. The knowledge of defect injecting methods and processes enable the defect prevention.

Implementation of defect prevention and detection strategies in an organization leads to the development of defect-free product. Identification of defect at the deployment stage or even at the later stages of development is highly expensive. Defect-free product has a direct and strong impact on the time, cost, and quality of the deliverables (Colligan 1997). It reduces support cost, programming cost, development time, and competitive advantage. Figure 2. depicts the cost quality analysis which indicates the impact of defects with respect to time and cost.

Insert Figure 2. about here

Author, (Jones 2008) states that defect potential and defect removal efficiency as the most critical software quality measurement in industry. Defect potential is the number of defects that occurs during the development of software applications. He further suggests the use of function points for the measurement of defect potentials, since defects occur at all phases of development.

Defects also occur due to the complexity of the project. Software is complex by nature. Complexity of software further builds up due to the difficulty in managing the development process. Thus, process and people contribute dealing with the complexity of the project. The development team requires effective coordination and co-operation among the team members to realize effective defect management.

Software industry follows several defect management strategies to address defect-associated issues. Two important approaches of effective defect management are quality control and quality assurance activities. Testing is a quality control activity that addresses defects. It can only show presence of defects and corrects them. Inspection is one of the most effective formal evaluation techniques of quality assurance. It detects most of the static defects at the early stages of software development and close to its origin. Inspection is both defect detection and prevention activity. For three decades, inspection has been the most mature, valuable, and competent technique in the challenging area of defect management (Fagan 2002; Tian 2005).

Inspection is carried out at every phase of the software development in order to uncover the maximum number of defects. Inspections initially reviewed by the individuals are later opened up for discussions amid the team members in formal technical meetings. The identified defect goes to the concerned developers for its refinement. Feedback mechanism facilitates developing team and management to identify and remove the defects along with fault processes [96].

Inspection examines the deliverable before implementation while testing examines the implemented deliverable. Therefore, inspection activities detect most of the static defects and thereby reduce test effort and cost of rework to fix those static defects. Thus, for an organization, investment on inspection at all phases of software development has a greater impact on Return on Investment (ROI). An informal attitude towards inspection leads to failures of acceptance test, which are highly undesirable.

Authors, (Kollanus and Koskinen 2007) present a survey of research work conducted in the area of inspection from 1991 to 2005. They recommend emphasizing on the implementation of inspection and its improvement as the most beneficial research area. They state that areas like reading techniques, evaluation of significance of inspection meetings, estimation of errors, peer review process are the most popular areas of research. They further express that the research in the direction of the introduction of novel inspection methods is very less.

Regardless of the significance of inspection, software industries consider inspection to be an activity, which is sporadically stringent and habitually overlooked. Its negligence adds to the accumulation of residual defects and demands more test effort to achieve effective defect management.

Our deep investigation of existing inspection technique in leading small and medium scale software industries shows their capability to deliver maximum of 96% defect-free product. The rationale for this is the lack of accurate perception of inspection performance, which is yet to be, achieved (Nair and Suma 2010a). The evaluation of inspection process was conventional with lesser theoretical analysis and comprehension of intricacies of its effect on ultimate productivity in industry.

The above investigation and claims indicate a strong need to enhance the existing inspection technique to achieve nearly zero-defect product. The study calls for enhancement in the existing inspection technique (process) and accentuate upon inspection team (people) to deploy nearly zero-defect product (effective defect management).

Metrics are numerical values that quantify the process, people and the product. They define, measure, manage, monitor and improve the effectiveness of the process and the product (Rico 2004). Accuracy in planning the inspection process improves with numerical estimations. Thus, inspection metrics improves the defect management (Borysowich 2006).

In order to have an appropriate measurement of the inspection process, the duo metric, the process metric, Depth of Inspection (DI) and people metric, Inspection Performance Metric (IPM) are introduced. Parameters such as number of inspectors, inspection time, inspection preparation time, experience level of inspector, skill of inspector in addition to complexity of the project modulates the value of DI and IPM (Nair and Suma 2010a). The introduction of pair metrics, DI and IPM into the process and people can yield valuable information of the company with regard to the inspection. It can further effectively reduce the overheads of software production that originates due to defects.

Our studies have indicated that the use of DI and IPM enables the manager within the software community to identify and compare the level of inspection and the inspection effort performed in various projects. They further enable one to estimate the quality levels (Nair and Suma 2010a). Introduction of the aforementioned metrics further help to quantify the ability of inspection process to capture defects within the constraints of parameters affecting inspection. The IPM could be organized as a performance-benchmarking tool for the projects in order to improve the in-house defect management process in software industry. DI and IPM will further pave the way for stakeholders to have a deep visibility into the process and thereby to justify the developmental cost. Implementation of these metrics reflects a continual process of improvement and success level of the organization.

This article elucidates a method to estimate the properties of a software company using the DI metric at the process level and using the IPM for estimating the properties of a team. The mathematical models presented here are capable of estimating a set of characteristic coefficients of the software company or team in such a way that these coefficients would enable later to calculate any required parameters, if the desired DI and IPM values are known. These parameters could be inspection time, preparation time, number of inspectors, experience level of inspectors and complexity of the project using function point analysis in a logarithmic scale.

## 2. RATIONALE FOR A MATHEMATICAL MODEL

A mathematical model is an abstract representation of the system using mathematical language. This model enables one to describe the properties of a system by providing an accurate reasoning for the observed properties.

Authors, (Woodcock and Loomes 1998) recommend the application of mathematical based evidences to real problems on an industrial scale. They express that mathematics is the only way through which complexities can be detected and resolved. Author, (Reiter1995) states that mathematical modeling supports two objectives namely i) to prove the truth of the research and ii) to support the research work through a sequence of mathematical relations. Author, (Henderson-Sellers 1996) feels that validity of metrics can be established through mathematical models. Author, (Riguzzi 1996) emphasizes that mathematical proof enables one to describe the process by assigning numbers or symbols to attributes such as processes, product and resources of real world entities. According to him, such an assignment preserves intuitive and empirical observations about the attributes. He thus expresses that mathematical model specifies relation between theory and empirical observations.

According to Goodman (2004), estimation of project through metrics is one of the best practices for a successful IT management. He states that estimation acts as the foundation for the work to progress and enables management within the software community to plan for the risk management.

Strategic investigations of empirical projects indicate that industry uses quality metrics to quantify defect profile such as defect removal efficiency (DRE) metric (Kan 2003; Wiegers 2007; Garmus and Harron 2006).

The existing metrics in industry, however, deals with latent defects and not on development of defect-free deliverables. There is a lack of awareness of a proactive process to quantify the inspection process.

## 3. RESEARCH METHODOLOGY

Our empirical study examines various projects developed in several leading service-based and product-based software industries. The objective of the research is to study deeply the inspection process, based on different categories of projects in a viable ensemble.

Our scope of investigation includes projects developed from the year 2000 onwards to 2009. As a mode of quality measurement unit, we consider three categories of projects, which are popular in the software houses. These categories include -- small project, which require less than 1000 person-hours for development, medium project, which need 1000 to 5000 person-hour of development time, and large project, which demands more than 5000 person-hour of total development time.

Function points are measurement units of the project. They are used as the comparison factor between projects for the quantitative analysis. The points of measurement include data functions and transaction functions. Data functions are internal logical files and external interface files. Transaction functions include external inputs, external outputs and external inquiries (Garmus and Harron 2006; Longstreet 1992).

Small projects consist of nearly less than 200 function points and less than 50 major requirements excluding trivial requirements such as user controllable interface. Medium projects consist of nearly less than 1000 function points and 50 to 100 major requirements excluding trivial requirements. Large projects contain more than 1000 function points and major requirements being greater than 100 in number excluding trivial requirements.

The base of this supposition is on the domain assumption that the implementation of one function point requires nearly five person-hours of development.

All the sampled projects are similar projects in terms of technology, environment, and programming language used. The work focuses on the three major phases of software development, namely, requirements analysis, design, and implementation phase.

## 4. INTRODUCTION OF DEPTH OF INSPECTION (DI), THE PROCESS METRIC

An operational definition for depth of inspection DI for a project is the number of defects captured by the inspection process divided by the total number of defects captured by both inspection and testing activities.

DI = Number of defects captured by inspection process (Ni)/Total number of defects captured by both inspection and testing (Td)

$$DI = Ni / Td \qquad (1)$$

Inclusion of the metric indicates a refined development process. DI metric measures the effectiveness of inspection. It is also a defect preventing and quality metric. DI computation for the projects enables the company to visualize its process maturity, and thereby is a lesson learned for future projects. DI emphasizes maximum defect capturing ability of the inspection process with the intention of reducing test effort. It neither encourages nor emphasizes the existence of latent defects. DI value enables one to fix the expected quality of software (Suma and Nair 2010c).

## 5. INTRODUCTION OF INSPECTION PERFORMANCE METRIC (IPM), THE PEOPLE METRIC

Operational definition for inspection performance analysis introduces the Inspection Performance Metric (IPM) (Nair and Suma 2010a).

**Inspection performance metric (IPM) = Number of defects captured by inspection process (Ni)/Inspection effort (IE)**

Where inspection effort (IE) = Total number of inspectors (N) * Total amount of inspection time (T)

And total amount of inspection time (T) = Actual inspection time (It) + Preparation time (Pt)

$$\textbf{IPM = Ni / IE} \qquad (2)$$

Where $IE = N \times T$
And $T = It + Pt$

The above metric is for a team and not for an individual.

### 5. 1. EMPIRICAL SUBSTANTIATION

Table 1., Table 2., Table 3. depicts the sampled projects from the leading service-based and product-based software industries. They illustrate the time, defect, and inspection-associated information at the three major phases of software development for small, medium and large category of projects respectively.

Insert Table 1. about here

Insert Table 2. about here

Insert Table 3. about here

### 5.2 EMPIRICAL OBSERVATIONS

Aforementioned information enables one to capture the functional properties and dependencies of variables contributing the inspection effectiveness. DI value is considered to be remaining between 0 and 1 and it takes up discrete values for the estimation purpose.

Our approach of mathematical modeling for the implementation of the duo metrics uses Multiple Linear Regression (MLR) scheme for the prediction of DI and IPM values. The rationale for the choice of linear regression model is that DI and IPM, which are the response variables (y), depend on more than one independent x variables. The model used above is first order MLR as the maximum power of the variables in the model is 1 (Montgomery 2006). The application method is found to be rewarding through field data and    is discussed in detail along with the methods and mode of usage.

## 5.3 ESTIMATION OF DI

The model consists of estimation of dependent variable DI through the four independent coefficients namely: $\beta_0$, $\beta_1$, $\beta_2$, $\beta_3$ and $\beta_4$ representing the process coefficients, which are estimated through modeling using the data from project information base. These coefficients are partial regression coefficients as they let the impact of other variables on the dependent variable. DI is evaluated using multiple linear regression models, whose details can be seen in Nair and Suma, 2010b.

## 5.4 ESTIMATION OF IPM

The similar approach is followed in developing a mathematical model for the estimation of people metric IPM (y) through the five independent observable coefficients characterizing a specific company or a team. The IPM is evaluated using the multiple linear regression models using the following equation

$$Y = \beta_0 + \beta_1 x_1 + \beta_2 x_2 + \beta_3 x_3 + \beta_4 x_4 + \beta_5 x_5 \tag{3}$$

The coefficients $\beta_0$, $\beta_1$, $\beta_2$, $\beta_3$, $\beta_4$ and $\beta_5$ represent the team coefficients, which are partial regression coefficients, estimated through modeling using the same set of data from project information base. The parameters are

$x_1$ = Inspection time
$x_2$ = Inspection Preparation time
$x_3$ = Number of inspectors
$x_4$ = Experience level of inspectors
$x_5$ = Complexity of the project using function point analysis in a logarithmic scale where $x_5$ is calculated as

$$x_{n5} = \frac{log10(function\ point)}{4} \tag{4}$$

Experience of inspectors is an influencing parameter in defect detection. Established projects require lesser time in elicitation of requirement than innovative projects. Hence innovative projects demand experienced inspectors. Integrated projects need more developmental time in design phase than other life cycle phases. Such projects demand experienced inspectors at design phase. An inspector who has examined design deliverables for minimum three projects is preferred for inspecting high-level design and low-level design. Experience level of inspectors can be considered in three categories. They are i) novice inspectors (up to 2 years) ii) average experienced inspectors (2 to 4 Years) and iii) largely experienced inspectors (above 4) (Nair and

Suma 2010a). In this model, an inspector is approved on the basis that he has achieved flat maturity level of carrying out inspection work rather than he is on a learning curve.

Logarithmic scales are used as the application consists of wide ranges of values. By applying the properties of natural logarithm, the parameter $x_5$, which is the complexity of the project, is realized within the range of 0 to 1 using equation numbered (1). Thus, in this mathematical modeling of IPM, complexity is taken as a variation from 0 to 1, and 1 being the complexity for a project consisting of 10000 function points.

### 5.5 IPM MODEL WITH PROJECT PARAMETERS

From the empirical observations made over several projects, equation (3) takes up the form of system of n equations as given below

$$IPM_1 = \beta_0 + \beta_1 x_{11} + \beta_2 x_{12} + \beta_3 x_{13} + \beta_4 x_{14} + \beta_5 x_{15}$$
$$IPM_2 = \beta_0 + \beta_1 x_{21} + \beta_2 x_{22} + \beta_3 x_{23} + \beta_4 x_{24} + \beta_5 x_{25}$$
$$\dots\dots\dots\dots$$
$$\dots\dots\dots\dots$$
$$IPM_i = \beta_0 + \beta_1 x_{i1} + \beta_2 x_{i2} + \beta_3 x_{i3} + \beta_4 x_{i4} + \beta_5 x_{i5}$$
$$IPM_n = \beta_0 + \beta_1 x_{n1} + \beta_2 x_{n2} + \beta_3 x_{n3} + \beta_4 x_{n4} + \beta_5 x_{n5}$$

The matrix notation for the aforementioned system of n equations is

$$[IPM] = [X][\beta] \tag{5}$$

where $[IPM] = [Parameters] \times [TeamCoefficients]$

$$IPM = \begin{bmatrix} IPM_1 \\ IPM_2 \\ IPM_3 \\ IPM_4 \\ - \\ - \\ IPM_n \end{bmatrix} \quad X = \begin{bmatrix} 1 & x_{11} & x_{12} & x_{13} & x_{14} & x_{15} \\ 1 & x_{21} & x_{22} & x_{23} & x_{24} & x_{25} \\ 1 & x_{31} & x_{32} & x_{33} & x_{34} & x_{35} \\ 1 & x_{41} & x_{42} & x_{43} & x_{34} & x_{45} \\ - & - & - & - & - \\ - & - & - & - & - \\ 1 & x_{n1} & x_{n2} & x_{n3} & x_{n4} & x_{n5} \end{bmatrix} \quad \beta = \begin{bmatrix} \beta_0 \\ \beta_1 \\ \beta_2 \\ \beta_3 \\ \beta_4 \\ \beta_5 \end{bmatrix}$$

Matrix IPM contains information about IPM value for n observations. Matrix X is parameter matrix, which contains information about all the IPM influencing parameters for which the observations are obtained. β matrix is the team coefficients matrix.

The team coefficients β0, β1, β2, β3, β4 and β5 are evaluated by using the concept of least squares. Least Square technique is the most popular technique applied to regression models in order to find the best fitting curve for independent x variable and its dependent y variable. Least square technique provides the best fit of the data points by computing the sum of squares of the differences of all the input data points such as $S_{xx}$, $S_{yy}$ and $S_{xy}$.

$$S_{xx} = \sum x^2 - \frac{(\sum x)^2}{n} \tag{6}$$

$$S_{yy} = \sum y^2 - \frac{(\sum y)^2}{n} \tag{7}$$

$$S_{xy} = \sum xy - \frac{(\sum x)(\sum y)}{n} \tag{8}$$

This technique reduces the error to zero [139]. To evaluate the team coefficients, the normal estimation equation will be:

$$\sum_{i=1}^{n} y_i = n\beta_0 + \beta_1 \sum_{i=1}^{n} x_{i1} + \beta_2 \sum_{i=1}^{n} x_{2i} + \beta_3 \sum_{i=1}^{n} x_{3i} + \beta_4 \sum_{i=1}^{n} x_{4i} + \beta_5 \sum_{i=1}^{n} x_{5i}$$

Where n= number of observations and i=1 to n

$$\sum_{i=1}^{n} x_{1i} y_i = \beta_0 \sum_{i=1}^{n} x_{1i} + \beta_1 \sum_{i=1}^{n} x_{1i}^2 + \beta_2 \sum_{i=1}^{n} x_{1i} x_{2i} + \beta_3 \sum_{i=1}^{n} x_{1i} x_{3i} + \beta_4 \sum_{i=1}^{n} x_{1i} x_{4i} + \beta_5 \sum_{i=1}^{n} x_{1i} x_{5i}$$

$$\sum_{i=1}^{n} x_{2i} y_i = \beta_0 \sum_{i=1}^{n} x_{2i} + \beta_1 \sum_{i=1}^{n} x_{2i} x_{1i} + \beta_2 \sum_{i=1}^{n} x_{2i}^2 + \beta_3 \sum_{i=1}^{n} x_{2i} x_{3i} + \beta_4 \sum_{i=1}^{n} x_{2i} x_{4i} + \beta_5 \sum_{i=1}^{n} x_{2i} x_{5i}$$

$$\sum_{i=1}^{n} x_{3i} y_i = \beta_0 \sum_{i=1}^{n} x_{3i} + \beta_1 \sum_{i=1}^{n} x_{3i} x_{1i} + \beta_2 \sum_{i=1}^{n} x_{3i} x_{2i} + \beta_3 \sum_{i=1}^{n} x_{3i}^2 + \beta_4 \sum_{i=1}^{n} x_{3i} x_{4i} + \beta_5 \sum_{i=1}^{n} x_{3i} x_{5i}$$

$$\sum_{i=1}^{n} x_{4i} y_i = \beta_0 \sum_{i=1}^{n} x_{4i} + \beta_1 \sum_{i=1}^{n} x_{4i} x_{1i} + \beta_2 \sum_{i=1}^{n} x_{4i} x_{2i} + \beta_3 \sum_{i=1}^{n} x_{4i} x_{3i}$$
$$+ \beta_4 \sum_{i=1}^{n} x_{4i}^{2} + \beta_5 \sum_{i=1}^{n} x_{4i} x_{5i}$$

$$\sum_{i=1}^{n} x_{5i} y_i = \beta_0 \sum_{i=1}^{n} x_{5i} + \beta_1 \sum_{i=1}^{n} x_{5i} x_{1i} + \beta_2 \sum_{i=1}^{n} x_{5i} x_{2i} + \beta_3 \sum_{i=1}^{n} x_{5i} x_{3i}$$
$$+ \beta_4 \sum_{i=1}^{n} x_{5i} x_{4i} + \beta_5 \sum_{i=1}^{n} x_{5i}^{2}$$

Table 4. shows the n coefficients for three categories of projects at three main phases of software development. It depicts the phase-wise team coefficients, which were evaluated by considering the empirical projects from different companies.

Insert Table 4. about here

## 6. APPLICATION OF MODEL

IPM is modeled based on the influencing variables, which are carefully selected across size of projects and complexity of projects in the industry. It was observed that team coefficients estimated from the first phase of empirical evaluation gives values of expected IPM within 10 percent variation from the real values observed in software house.

Table 5. shows a sample of three projects taken in each category from industry. The three sampled projects, which are shown for the verification purpose, can be used to demonstrate the use of IPM. The table depicts the IPM value computed using the aforementioned team coefficients (IPM$_{tc}$) and IPM obtained from the industry based on defect count (IPM$_{dc}$). Figure 1., Figure 2., and Figure 3. depict the comparative results of IPM values based on defect count verses the model produced values at the three major phases of software development..

Insert Table 5. about here

Insert Figure 3. about here
Insert Figure 4. about here
Insert Figure 5. about here

The Figure 2. indicates Project 2 displaying a variation between IPM$_{dc}$ and IPM$_{tc}$ values which might have occurred due to decrease in design time scheduled in the project.

## 6.1 RESULTS

The estimation of this characteristic vector can enable the company and the outsourcer to analyze the inspection process and the degree of level of inspection process they want. Having finalized the IPM that a company should achieve, it can tune the number of persons doing inspection, the experience of each person, the time to be spent by each person essentially to achieve the desired quality level of IPM, using the coefficients estimated from earlier performance.

$R^2$ analysis of the input variables requires a wide spectrum of projects and the process is under way. A subsequent study report analyzing the properties of variables formulating the performance parameters like IPM will be the next step of analysis activity.

The sensitivity analysis of the chosen input variables is evaluated using the obtained team coefficients on the verification projects. Table 6., Table 7. and Table 8. depicts the sensitivity analysis of the chosen variables.

Insert Table 6. about here
Insert Table 7. about here
Insert Table 8. about here

The sensitivity analysis of the input variable indicates the variation of IPM value with a variation of the input variable values amounting to 10% as an example for the obtained team coefficients. This further implies that IPM can be tuned for the best fit based on the apt input variable values.

IPM is not a normalized index. The empirical data collected from a software company based on the assumptions as explained in the research methodology section helps to calculate the current people performance in terms of inspection performance metric. Using a set of real world values obtained from projects, it is possible to calculate a set of coefficients which governs the properties of the team which brings effective performance. The mathematical system is based on regression analysis and using these coefficients, it is possible to access different properties of the team for that particular company.

## 6.2 DISCUSSION

Management of the software community can reduce test effort by choosing an appropriate value for DI and IPM. This is because effective inspection can capture maximum defects. This technique will further aid the Personal Software Process (PSP) and Team Software Process (TSP) in an incredible way by numerically tabulating the performance and giving feedback to the members and to the team as a whole.

Author, (Humphrey 2004b) states that Personal software process (PSP) focuses on the quality of the software, which is achievable through defect management. Hence, he feels that product quality enhances with process quality. He suggests that review process captures more defects than testing. The aim of team software process (TSP) is to enable developing team to develop quality software in a disciplined work environment. This involves skilled, cohesive, and interdependent team members whose objective is to meet the expected schedule. Further, TSP emphasizes upon the review and prevention of defects through effective team collaboration (Humphrey 2004a, Humphrey 2000).

Introduction of DI, which is a process metric and IPM which is a people metric enables to quantify the ability of inspection process to capture defects within the constraints of parameters affecting inspection (Nair and Suma 2010a).

The CMM level 4, "Managed", quantitatively analyzes and controls both the software process and products. Therefore, characteristic coefficients of Software Company can vary in an admissible range of 1 to 10% within the framework of the hypothesis put forward in the research methodology section. It is inevitable to note that the ranges of values for acceptable performance differs from one type of project to another like innovative projects, legacy projects etc.

Author, James expresses that implementation of PSP and TSP in CMMI enhances the performance of team and the process. Further, TSP is found to support the key practices of the CMMI (Tamura Shurei 2009).

Further, stakeholders can have a hold on the process and team visibility due to the introduction of the pair metrics, which provides the transparency to the operations of the company. In the field of outsourcing, the visibility or expectations of performance and maturity of a company was highly intuitive till now based on skill of the assessment of the outsourcing agency. However, by using the aforementioned methodology, one can calculate the inspection performance of the company based on the previous inspection pattern unambiguously leading to cost saving through higher quality of codes and reduced test effort.

Managers get the added advantage of monitoring team performance project after project in a convincing way using numerical estimations through characteristic coefficients of the team or the company.

The DI and IPM value can now be either estimated based on defect count from the shop floor or can be predicted through the process coefficients and team coefficients, which were empirically, evaluated using a large sample of projects. Once the coefficients are stabilized, it is possible to predict the achievable DI and IPM through our model, without depending on defect count. It implies that the managers can have the ability to finalize the $x_1$, $x_2$, $x_3$, $x_4$ parameters while planning the inspection process to achieve a particular DI. Having finalized the IPM that a company should achieve, it can tune the number of persons doing inspection, the experience of each person, the time to be spent by each person essentially to achieve the desired quality level of IPM.

Since DI and IPM are directly affecting the defect management, development of 99 percent defect-free product is possible by choosing appropriate values of parameters influencing DI and IPM. Hence, the implementation of our enhanced approach will be a boon in effective process management, as it has become easy to carry out the quality control through effective defect management.

## 7. CONCLUSION

Effective defect management is one of the significant activities in software industry. Inspection claims to be one of the successful approaches toward effective defect management.

An empirical study of various projects across leading service-based and product-based software industries indicated the need for the introduction of a metric based process to quantify the inspection. The inspection process metric, The Depth of Inspection (DI) enables one to have a deep visibility of the process and it further helps to justify the developmental cost.

With the introduction of people metric, Inspection Performance Metric (IPM), the managers within the software community can analyze the desirable band of inspection effort required to achieve the desirable level of DI. A careful investigation of several projects facilitated the evaluation of a set of team coefficients, which could be used to predict the IPM values.

For the verification purpose, a comparative analysis of data is shown. The observed IPM value for a project and the IPM value produced by the model are found to be strongly matching. Hence, the model offers a predictive capability for IPM through team coefficients which could be used for tuning purposes for desired quality and cost levels even before starting projects on floor.

The estimation of pair metric using characteristic coefficients enables the company to evaluate the test effort, productivity and quality level of the company. Further, the aforementioned approach is a solution to several issues related to management, stakeholders, outsourcing agents and customers. In addition, the pair metric and characteristic coefficients support the strategy of PSP and TSP.

**ACKNOWLEDGMENTS**

The authors would like to acknowledge all the people from various software houses for their valuable discussions and immense help rendered in carrying out this work.

McHale D. James. Why Implement CMMI with PSP/TSP?, A PSP/TSP Approach to CMMI Transition, A PSP/TSP Approach to CMMI Transition,
Tamura Shurei. 2009. *Integrating CMMI and TSP/PSP: Using TSP Data to Create Process Performance Models.* Technical Note, CMU/SEI-2009-TN-033, November 2009
ZHANG Lina, LI Ya
[Software Process Improvement for Small Organizations Based on CMMI/TSP/PSP].

Table 1. Inspection and defect profile of small project at the three major phases of software development

| Phase | Project | P1 | P2 | P3 | P4 | P5 | P6 |
|---|---|---|---|---|---|---|---|
| | Total Project hours (person-hour) | 250 | 263 | 300 | 507 | 800 | 869 |
| Requirements Analysis | Total defects | 30 | 35 | 46 | 77 | 64 | 58 |
| | Defects captured through inspection | 16 | 17 | 31 | 40 | 31 | 19 |
| | Defects captured through testing | 14 | 18 | 15 | 37 | 33 | 39 |
| | Inspection time | 3 | 3 | 4 | 6 | 24 | 6 |
| | Preparation time | 0.5 | 0.15 | 0.5 | 1 | 2 | 1 |
| | Number of inspectors | 3 | 3 | 3 | 3 | 3 | 3 |
| | Experience level of inspectors (years) | 1 | 1 | 1 | 2 | 2 | 5 |
| Design | Total defects | 10 | 8 | 13 | 26 | 25 | 38 |
| | Defects captured through inspection | 5 | 3 | 6 | 14 | 13 | 16 |
| | Defects captured through testing | 5 | 5 | 7 | 12 | 12 | 22 |
| | Inspection time | 6 | 4 | 5 | 11 | 30 | 16 |
| | Preparation time | 1 | 0.5 | 1 | 1 | 3 | 2 |
| | Number of inspectors | 3 | 3 | 4 | 4 | 4 | 3 |
| | Experience level of inspectors (years) | 2 | 2 | 2 | 3 | 2 | 5 |
| Implementation | Total defects | 8 | 14 | 16 | 17 | 36 | 19 |
| | Defects captured through inspection | 4 | 8 | 7 | 9 | 16 | 7 |
| | Defects captured through testing | 4 | 6 | 9 | 8 | 20 | 12 |
| | Inspection time | 3 | 3 | 3 | 3 | 25 | 3 |
| | Preparation time | 1 | 1.5 | 2 | 2 | 2.5 | 2 |
| | Number of inspectors | 3 | 3 | 3 | 3 | 3 | 3 |
| | Experience level of inspectors (years) | 2 | 2 | 2 | 3 | 3 | 5 |

Table 2. Inspection and defect profile of medium project at the three major phases of software development

| Phase | Project | P7 | P8 | P9 | P10 | P11 | P12 |
|---|---|---|---|---|---|---|---|
| | **Total Project hours (person-hour)** | 1806 | 2110 | 3000 | 4248 | 4586 | 4644 |
| **Requirements analysis** | Total defects | 58 | 139 | 130 | 175 | 200 | 150 |
| | Defects captured through inspection | 28 | 69 | 60 | 80 | 77 | 40 |
| | Defects captured through testing | 30 | 70 | 70 | 95 | 123 | 110 |
| | Inspection time | 7 | 48 | 101 | 107 | 200 | 36 |
| | Preparation time | 2 | 7 | 12 | 15 | 16 | 3 |
| | Number of inspectors | 3 | 4 | 4 | 5 | 3 | 3 |
| | Experience level of inspectors (years) | 5 | 3 | 4 | 5 | 2 | 5 |
| **Design** | Total defects | 38 | 55 | 42 | 70 | 75 | 70 |
| | Defects captured through inspection | 19 | 24 | 22 | 34 | 33 | 28 |
| | Defects captured through testing | 19 | 31 | 20 | 36 | 42 | 42 |
| | Inspection time | 20 | 48 | 156 | 143 | 128 | 16 |
| | Preparation time | 2 | 7 | 24 | 25 | 25 | 2 |
| | Number of inspectors | 3 | 5 | 5 | 3 | 4 | 3 |
| | Experience level of inspectors (years) | 5 | 4 | 5 | 6 | 6 | 6 |
| **Implementation** | Total defects | 38 | 36 | 50 | 47 | 53 | 15 |
| | Defects captured through inspection | 8 | 14 | 28 | 24 | 27 | 6 |
| | Defects captured through testing | 30 | 22 | 22 | 23 | 26 | 9 |
| | Inspection time | 42 | 95 | 80 | 105 | 91 | 16 |
| | Preparation time | 2 | 15 | 8 | 14 | 16 | 3 |
| | Number of inspectors | 3 | 5 | 5 | 5 | 5 | 3 |
| | Experience level of inspectors (years) | 5 | 5 | 4 | 6 | 6 | 5 |

Table3. Inspection and defect profile of large project at the three major phases of software development

| Phase | Project | P11 | P12 | P13 | P14 | P15 | P16 |
|---|---|---|---|---|---|---|---|
| | **Total Project hours (person-hour)** | 6944 | 7087 | 7416 | 8940 | 9220 | 12000 |
| **Requirements analysis** | Total defects | 254 | 400 | 320 | 450 | 375 | 410 |
| | Defects captured through inspection | 112 | 175 | 156 | 200 | 175 | 250 |
| | Defects captured through testing | 142 | 225 | 164 | 250 | 200 | 160 |
| | Inspection time | 281 | 225 | 235 | 234 | 250 | 450 |
| | Preparation time | 42 | 40 | 69 | 40 | 42.12 | 60 |
| | Number of inspectors | 7 | 4 | 3 | 4 | 5 | 5 |
| | Experience level of inspectors (years) | 7 | 6 | 3 | 6 | 3 | 6 |
| **Design** | Total defects | 120 | 175 | 150 | 200 | 182 | 200 |
| | Defects captured through inspection | 77 | 80 | 86 | 90 | 78 | 140 |
| | Defects captured through testing | 43 | 95 | 64 | 110 | 104 | 60 |
| | Inspection time | 156 | 100 | 116 | 250 | 264 | 480 |
| | Preparation time | 33 | 50 | 61 | 60 | 123 | 96 |
| | Number of inspectors | 4 | 4 | 3 | 4 | 6 | 5 |
| | Experience level of inspectors (years) | 6 | 6 | 4 | 6 | 4 | 6 |
| **Implementation** | Total defects | 67 | 120 | 70 | 150 | 98 | 115 |
| | Defects captured through inspection | 37 | 60 | 32 | 70 | 48 | 77 |
| | Defects captured through testing | 30 | 60 | 38 | 80 | 50 | 38 |
| | Inspection time | 156 | 100 | 116 | 250 | 264 | 750 |
| | Preparation time | 32 | 20 | 45 | 40 | 141 | 150 |
| | Number of inspectors | 3 | 4 | 6 | 4 | 4 | 5 |
| | Experience level of inspectors (years) | 7 | 6 | 4 | 6 | 5 | 4 |

Table 4. Phase-Wise Team Coefficients spanning the size complexity

| Phase | Team coefficients | Small project | Medium project | Large project |
|---|---|---|---|---|
| Requirements | $\beta_0$ | 0 | 4.4683 | 0.8698 |
| | $\beta_1$ | -379.8835 | -19.7096 | -1.3141 |
| | $\beta_2$ | -255.4827 | 108.3653 | -22.3141 |
| | $\beta_3$ | 23.5850 | -0.3135 | -0.1130 |
| | $\beta_4$ | -3.9719 | 0.0933 | 0.0030 |
| | $\beta_5$ | -23.2132 | -3.0703 | 0.9589 |
| Design | $\beta_0$ | 1.0748 | -3.2532 | 0.5904 |
| | $\beta_1$ | 8.3533 | 5.6946 | 1.2494 |
| | $\beta_2$ | -28.0972 | -227.1285 | 1.6164 |
| | $\beta_3$ | 0.6315 | 0.2357 | -0.0314 |
| | $\beta_4$ | 0.3499 | -0.1879 | 0.0609 |
| | $\beta_5$ | -8.0960 | 9.9915 | -0.8328 |
| Implementation | $\beta_0$ | 0 | 0.3796 | 13.7607 |
| | $\beta_1$ | -6.5943 | -4.9077 | -14.0781 |
| | $\beta_2$ | 40.3309 | 9.1943 | -14.0568 |
| | $\beta_3$ | -0.0797 | 0.0205 | -0.4536 |
| | $\beta_4$ | -0.1705 | -0.0499 | -0.5835 |
| | $\beta_5$ | 2.9149 | 0.6252 | -7.7391 |

Table 5. Results from the implementation of IPM in verification projects

| Phase | Project | P1 | P2 | P3 |
|---|---|---|---|---|
| | Project hours (person-hour) | 1000 | 3500 | 10600 |
| Requirements analysis | Req. time | 150 | 940 | 1590 |
| | Total defects | 46 | 115 | 200 |
| | Defects captured through inspection | 21 | 55 | 124 |
| | Defects captured through testing | 25 | 60 | 76 |
| | Inspection time | 16.5 | 94 | 223 |
| | Preparation time | 1.5 | 10 | 32 |
| | Number of inspectors | 3 | 4 | 5 |
| | Experience of inspectors (years) | 3 | 4 | 8 |
| | IPM (based on defect count) | 1.16 | 0.5288 | 0.4862 |
| | IPM (based on team coefficients) | 1.1437 | 0.5164 | 0.4959 |
| Design | Design time | 250 | 530 | 2650 |
| | Total defects | 54 | 60 | 150 |
| | Defects captured through inspection | 28 | 25 | 92 |
| | Defects captured through testing | 26 | 35 | 58 |
| | Inspection time | 27.5 | 64 | 345 |
| | Preparation time | 2.5 | 11 | 80 |
| | Number of inspectors | 4 | 4 | 5 |
| | Experience of inspectors (years) | 4 | 4 | 6 |
| | IPM (based on defect count) | 0.9333 | 0.3333 | 0.2164 |
| | IPM (based on team coefficients) | 0.9810 | 0.0716 | 0.3172 |
| Implementation | Implementation time | 400 | 1000 | 4240 |
| | Total defects | 25 | 45 | 225 |
| | Defects captured through inspection | 16 | 23 | 96 |
| | Defects captured through testing | 9 | 22 | 129 |
| | Inspection time | 44 | 100 | 509 |
| | Preparation time | 4 | 10 | 85 |
| | Number of inspectors | 4 | 4 | 5 |
| | Experience of inspectors (years) | 4 | 6 | 5 |
| | IPM (based on defect count) | 0.333 | 0.209 | 0.1616 |
| | IPM (based on team coefficients) | 0.3540 | 0.2081 | 0.1690 |

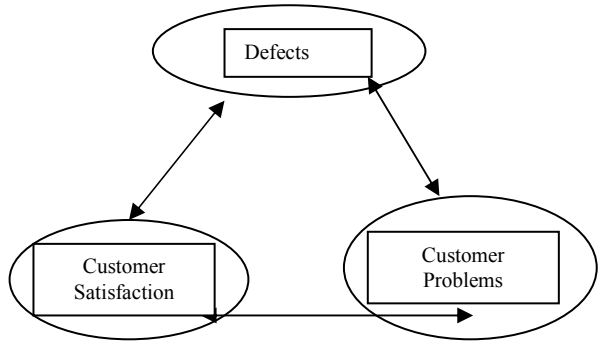

Figure 1. Interdependency of customer satisfaction and defects

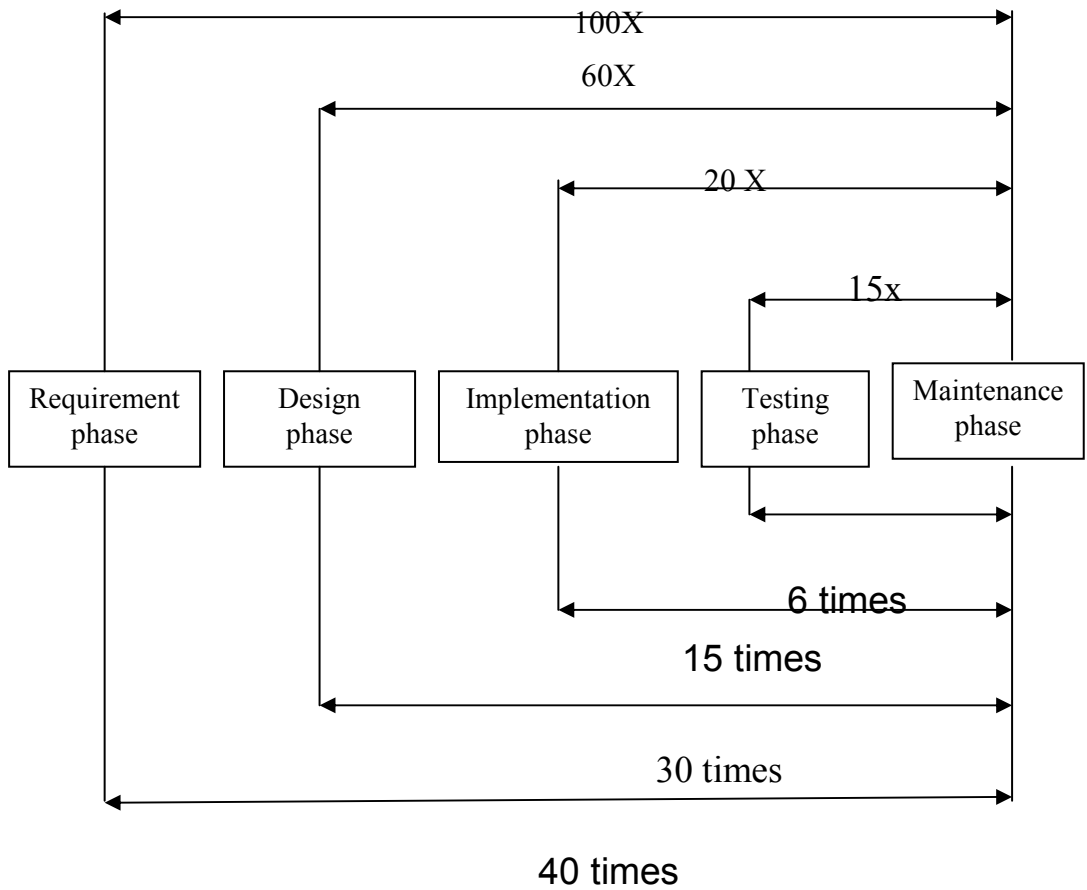

Relative time to fix software defects

Figure 2. Cost-Quality Analysis

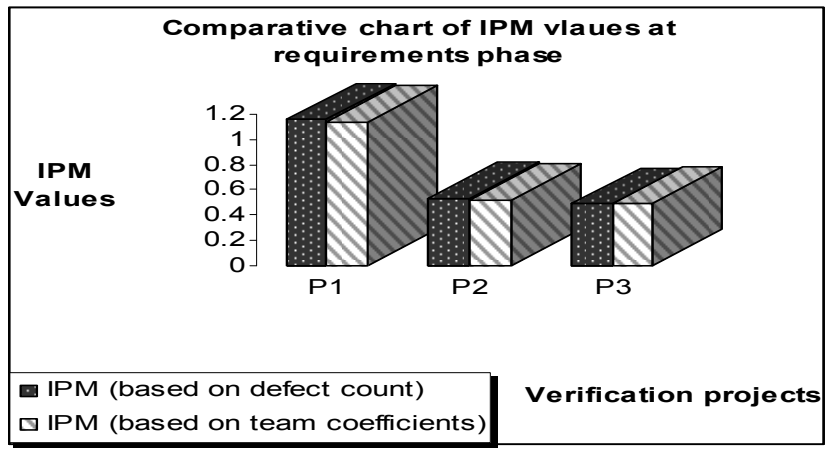

Figure 3. Comparative chart of IPM values at requirements phase of verification projects

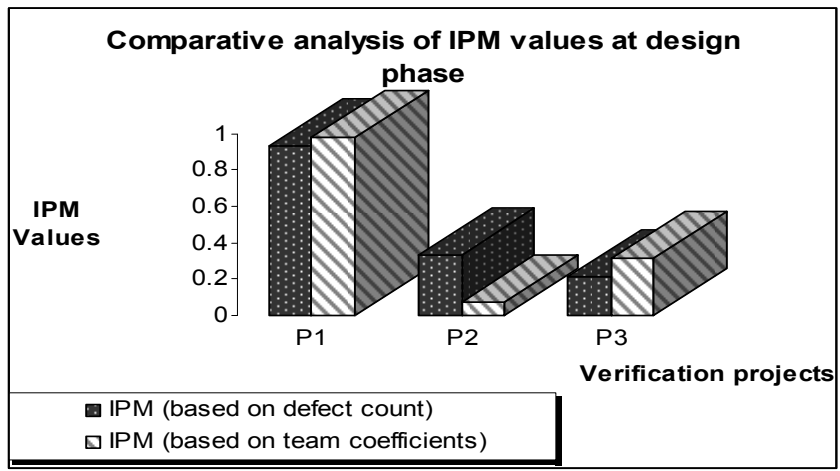

Figure 4. Comparative chart of IPM values at design phase of verification projects

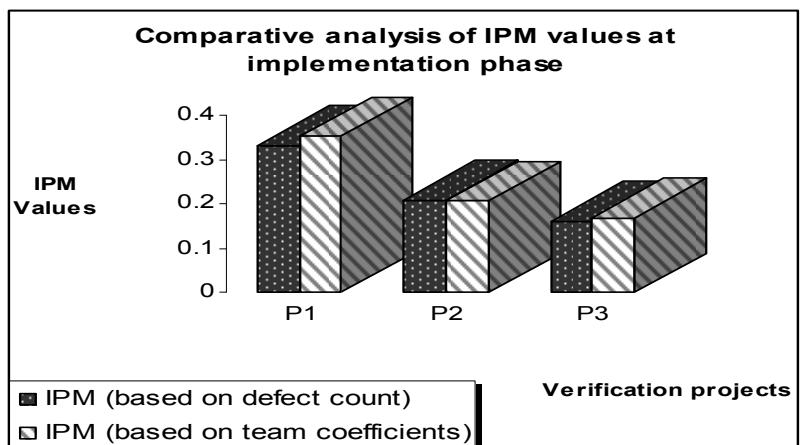

Figure 5. Comparative chart of IPM values at implementation phase of verification projects

Table 6. Sensitvity analysis of input vairables for the verification project P1

| Phase | Project | P1 | P1 | P1 | P1 | P1 | P1 | P1 | P1 | P1 |
|---|---|---|---|---|---|---|---|---|---|---|
| | Project hours (person-hour) | 1000 | 1000 | 1000 | 1000 | 1000 | 1000 | 1000 | 1000 | 1000 |
| Requirements analysis | Req. time | 150 | 150 | 150 | 150 | 150 | 150 | 150 | 150 | 150 |
| | Inspection time | 16.5 | 19.5 | 16.5 | 16.5 | 16.5 | 15 | 16.5 | 16.5 | 16.5 |
| | Preparation time | 1.5 | 1.5 | 4 | 1.5 | 1.5 | 1.5 | 0.5 | 1.5 | 1.5 |
| | Number of inspectors | 3 | 3 | 3 | 4 | 3 | 3 | 3 | 2 | 3 |
| | Experience of inspectors (years) | 3 | 3 | 3 | 3 | 5 | 3 | 3 | 3 | 1 |
| | IPM (based on inspection effort coefficients) | 1.1437 | -6.4481 | -1.4052 | 24.73495 | -6.7943 | 4.94855 | -9.0699 | -22.435 | 9.0936 |
| Design | Design time | 250 | 250 | 250 | 250 | 250 | 250 | 250 | 250 | 250 |
| | Inspection time | 27.5 | 32.5 | 27.5 | 27.5 | 27.5 | 20 | 27.5 | 27.5 | 27.5 |
| | Preparation time | 2.5 | 2.5 | 6.5 | 2.5 | 2.5 | 2.5 | 1 | 2.5 | 2.5 |
| | Number of inspectors | 4 | 4 | 4 | 5 | 4 | 4 | 4 | 2 | 4 |
| | Experience of inspectors (years) | 4 | 4 | 4 | 4 | 5 | 4 | 4 | 4 | 1 |
| | IPM (based on process coefficients) | 0.9811 | 1.1485 | 0.07 | 1.6129 | 1.33135 | 0.73085 | -0.1424 | -0.2816 | -0.0683 |
| Implementation | Implementation time | 400 | 400 | 400 | 400 | 400 | 400 | 400 | 400 | 400 |
| | Inspection time | 44 | 52 | 44 | 44 | 44 | 32 | 44 | 44 | 44 |
| | Preparation time | 4 | 4 | 10 | 4 | 4 | 4 | 1.5 | 4 | 4 |
| | Number of inspectors | 4 | 4 | 4 | 5 | 4 | 4 | 4 | 2 | 4 |
| | Experience of inspectors (years) | 3 | 3 | 3 | 3 | 5 | 3 | 3 | 3 | 1 |
| | IPM (based on process coefficients) | 0.355 | 0.2067 | 0.7419 | 0.2489 | 0.1681 | 0.5364 | 1.9518 | 0.498 | 0.8501 |

Table 7. Sensitvity analysis of input vairables for the verification project P2

| Phase | | | | | | | | | | |
|---|---|---|---|---|---|---|---|---|---|---|
| | Project hours (person-hour) | 3500 | 3500 | 3500 | 3500 | 3500 | 3500 | 3500 | 3500 | 3500 |
| Requirements analysis | Req. time | 940 | 940 | 940 | 940 | 940 | 940 | 940 | 940 | 940 |
| | Inspection time | 94 | 112 | 94 | 94 | 94 | 80 | 94 | 94 | 94 |
| | Preparation time | 10 | 10 | 22 | 10 | 10 | 10 | 4 | 10 | 10 |
| | Number of inspectors | 4 | 4 | 4 | 5 | 4 | 4 | 4 | 2 | 4 |
| | Experience of inspectors (years) | 4 | 4 | 4 | 4 | 6 | 4 | 4 | 4 | 2 |
| | IPM (based on inspection effort coefficients) | 0.5163 | 0.1232 | 1.5999 | 0.2034 | 0.704 | 0.3061 | 0.5757 | 0.167 | 0.4076 |
| Design | Design time | 530 | 530 | 530 | 530 | 530 | 530 | 530 | 530 | 530 |
| | Inspection time | 64 | 69 | 64 | 64 | 64 | 42 | 64 | 64 | 64 |
| | Preparation time | 11 | 11 | 21 | 11 | 11 | 11 | 4 | 11 | 11 |
| | Number of inspectors | 4 | 4 | 4 | 7 | 4 | 4 | 4 | 2 | 4 |
| | Experience of inspectors (years) | 4 | 4 | 4 | 4 | 7 | 4 | 4 | 4 | 5 |
| | IPM (based on process coefficients) | 0.1854 | -1.4513 | -1.2748 | 0.4005 | -0.906 | 0.1854 | 0.1854 | 0.1854 | 0.1854 |
| Implementation | Implementation time | 1000 | 1000 | 1000 | 1000 | 1000 | 1000 | 1000 | 1000 | 1000 |
| | Inspection time | 100 | 130 | 100 | 100 | 100 | 80 | 100 | 100 | 100 |
| | Preparation time | 10 | 10 | 26 | 10 | 10 | 10 | 4 | 10 | 10 |
| | Number of inspectors | 4 | 4 | 4 | 6 | 4 | 4 | 4 | 2 | 4 |
| | Experience of inspectors (years) | 6 | 6 | 6 | 6 | 8 | 6 | 6 | 6 | 2 |
| | IPM (based on process coefficients) | 0.2076 | 0.0607 | 0.2999 | 0.249 | 0.1082 | 0.3061 | 0.5757 | 0.167 | 0.4076 |

Table 8. Sensitvity analysis of input vairables for the verification project P3

| Phase | Project | P3 | P3 | P3 | P3 | P3 | P3 | P3 | P3 | P3 |
|---|---|---|---|---|---|---|---|---|---|---|
| | Project hours (person-hour) | 10600 | 10600 | 10600 | 10600 | 10600 | 10600 | 10600 | 10600 | 10600 |
| Requirements analysis | Req. time | 1590 | 1590 | 1590 | 1590 | 1590 | 1590 | 1590 | 1590 | 1590 |
| | Inspection time | 223 | 238.5 | 223 | 223 | 223 | 160 | 223 | 223 | 223 |
| | Preparation time | 32 | 32 | 95 | 32 | 32 | 32 | 16 | 32 | 32 |
| | Number of inspectors | 5 | 5 | 5 | 7 | 5 | 5 | 5 | 3 | 5 |
| | Experience of inspectors (years) | 8 | 8 | 8 | 8 | 9 | 8 | 8 | 8 | 3 |
| | IPM (based on inspection effort coefficients) | 0.4899 | 0.2565 | 0.0597 | 0.0568 | 0.2887 | 0.3222 | 0.7291 | 0.5088 | 0.2708 |
| Design | Design time | 2650 | 2650 | 2650 | 2650 | 2650 | 2650 | 2650 | 2650 | 2650 |
| | Inspection time | 345 | 398 | 345 | 345 | 345 | 212 | 345 | 345 | 345 |
| | Preparation time | 80 | 80 | 159 | 80 | 80 | 80 | 10.5 | 80 | 80 |
| | Number of inspectors | 5 | 5 | 5 | 7 | 5 | 5 | 5 | 3 | 5 |
| | Experience of inspectors (years) | 6 | 6 | 6 | 6 | 8 | 6 | 6 | 6 | 3 |
| | IPM (based on process coefficients) | 0.3135 | 0.2452 | 0.204 | 0.1574 | 0.342 | 0.1577 | 0.1879 | 0.283 | 0.0375 |
| Implementation | Implementation time | 4240 | 4240 | 4240 | 4240 | 4240 | 4240 | 4240 | 4240 | 4240 |
| | Inspection time | 509 | 552 | 509 | 509 | 509 | 340 | 509 | 509 | 509 |
| | Preparation time | 85 | 85 | 165.5 | 85 | 85 | 85 | 17 | 85 | 85 |
| | Number of inspectors | 5 | 5 | 5 | 7 | 5 | 5 | 5 | 3 | 5 |
| | Experience of inspectors (years) | 5 | 5 | 5 | 5 | 8 | 5 | 5 | 5 | 3 |
| | IPM (based on process coefficients) | 0.1691 | 0.0289 | 0.0291 | -0.2838 | -0.4137 | 0.7328 | 0.3102 | 1.0769 | 1.3367 |